\documentclass[reprint,amsmath,amssymb,aps]{revtex4-2}
\usepackage{blindtext}
\usepackage{CJK}
\usepackage{graphicx}
\usepackage{dcolumn}
\usepackage{bm}
\usepackage{lipsum}

\usepackage[usenames,dvipsnames]{xcolor}

\begin{document}

\title{Magnetochiral Properties of Spin Waves Existing in Nanotubes with Axial and Circumferential Magnetization}

\author{M. C. Giordano} \affiliation{\'Ecole Polytechnique F\'ed\'erale de Lausanne (EPFL), School of Engineering,  Institute of Materials, Laboratory of Nanoscale Magnetic Materials and Magnonics, 1015 Lausanne, Switzerland}
\author{M. Hamdi} \affiliation{\'Ecole Polytechnique F\'ed\'erale de Lausanne (EPFL), School of Engineering,  Institute of Materials, Laboratory of Nanoscale Magnetic Materials and Magnonics,  1015 Lausanne, Switzerland}
\author{A. Mucchietto} \affiliation{\'Ecole Polytechnique F\'ed\'erale de Lausanne (EPFL), School of Engineering,  Institute of Materials, Laboratory of Nanoscale Magnetic Materials and Magnonics, 1015 Lausanne, Switzerland}
\author{D. Grundler} \email[]{dirk.grundler@epfl.ch}  \affiliation{\'Ecole Polytechnique F\'ed\'erale de Lausanne (EPFL), School of Engineering,  Institute of Materials, Laboratory of Nanoscale Magnetic Materials and Magnonics,  1015 Lausanne, Switzerland} \affiliation{\'Ecole Polytechnique F\'ed\'erale de Lausanne, School of Engineering,  Institute of Electrical and Micro Engineering, 1015 Lausanne, Switzerland}

\vskip 0.25cm
\date{\today}

\begin{abstract}

We report experimental studies of spin-wave excitations in individual 22 nm thick Ni\textsubscript{80}Fe\textsubscript{20} nanotubes with diameters of about 150 nm by means of Brillouin light-scattering (BLS) spectroscopy. Irradiated by microwaves we resolve sets of discrete resonances in the center of nanotubes ranging from 2.5 to 12.5 GHz. Comparing to a recent theoretical work and micromagnetic simulations, we identify different characteristic eigenmodes depending on the axial, mixed or vortex configuration. The mixed and vortex states give rise to modes with helical phase profiles substantiating an unusual nature of modes attributed to non-reciprocal spin waves. Our findings provide microscopic insight into tubular spin-wave nanocavities and magnetochiral effects for 3D nanomagnonics.\end{abstract}



\maketitle

\begin{figure}[h!]
	\includegraphics[width=0.48\textwidth]{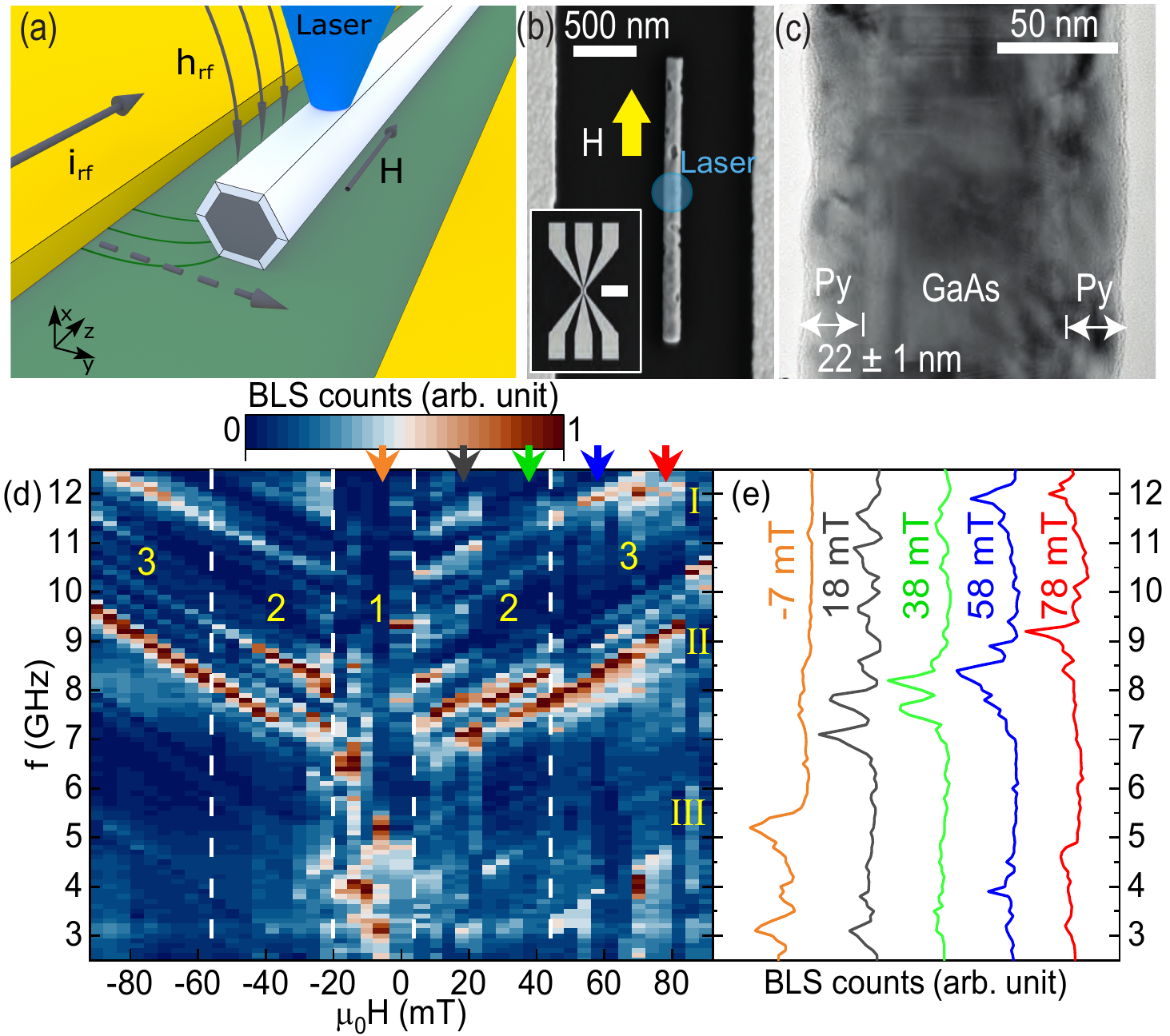}\centering
	\caption{(a) Sketched excitation-detection scheme based on a coplanar wave guide (CPW) and microfocus Brillouin light scattering microscopy. (b) Scanning electron microscopy image of the nanotube placed parallel to the CPW (inset: CPW at a smaller magnification: scale bar: 200 $\mu$m). (c) Transmission electron microscopy image of a Py nanotube on a GaAs core. Shell thickness: 22 $\pm$ 1 nm. (d) BLS spectra detected at room temperature at the NT center as a function of field $H$ applied along the NT axis. Colored arrows indicate fields $-7$, 18, 38, 58, and 78 mT at which we extract (e) line spectra. }\label{fig:experiment}
\end{figure}

\begin{figure}[t]
	\includegraphics[width=0.49\textwidth]{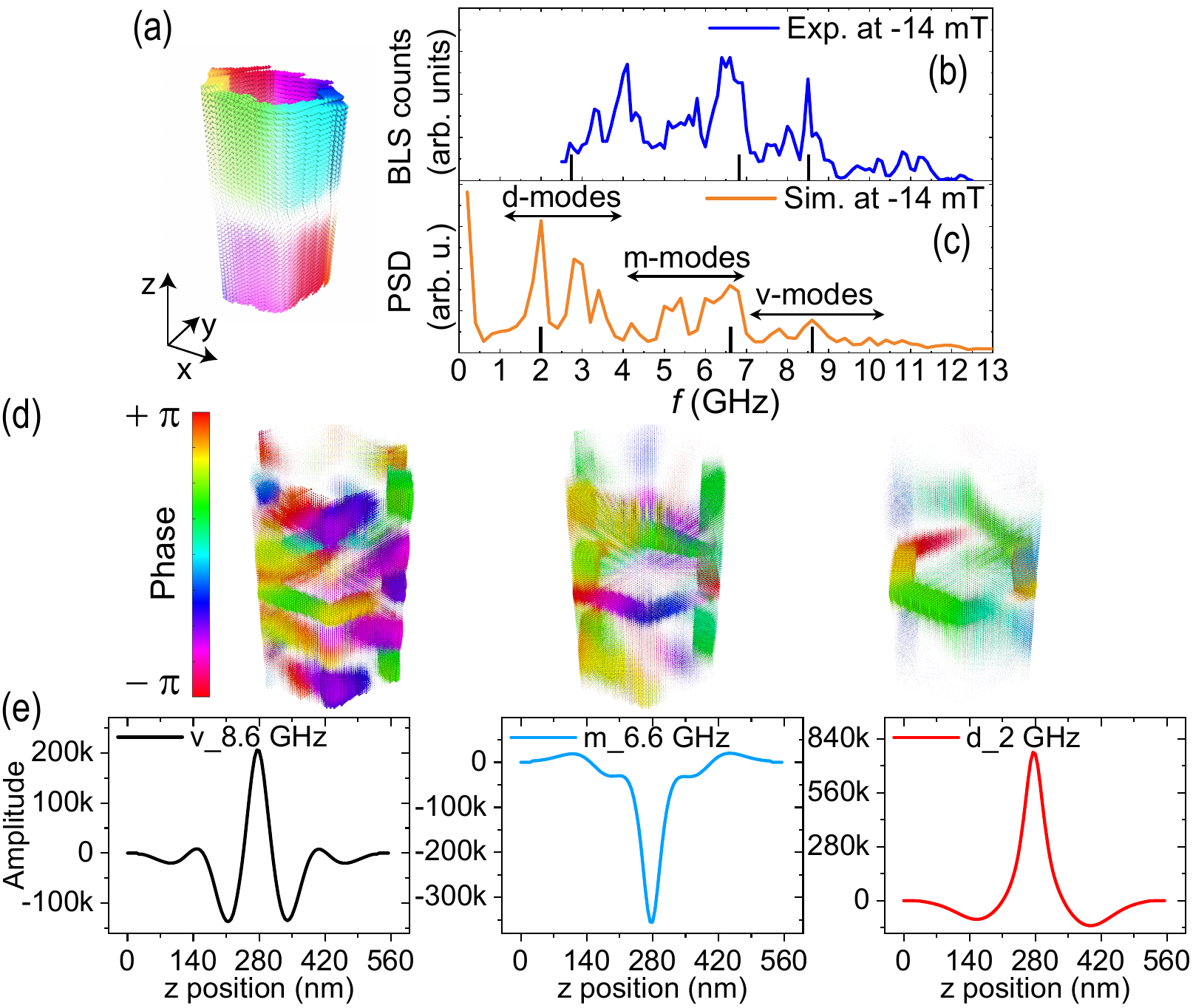}\centering
	\caption{(a) Simulated static magnetization profile of the NT at $\mu_{0}H_{z}$ = -14 mT. (b) BLS and (c) simulated spectrum at $\mu_{0}H_{z}$ = -14 mT. Eigenmodes in (c) are subdivided in three groups d, m and v. Black ticks refer to mode profiles and amplitudes displayed in (d) and (e), respectively. The size of the dots represent the amplitude. The color bar represents spin wave phase ranging from $-\pi$ to $+\pi$. (e) Spin-precessional amplitudes are plotted along \textbf{z} for a fixed position in the \textbf{x}-\textbf{y} plane of the NT.}\label{fig:dynamicSwitch}
\end{figure}

Advances in magnonics fosters new ideas for information processing based on reciprocal and non-reciprocal short-wave magnons \cite{Kruglyak2010,Vogt2014,Gubbiotti2019}. They set novel grounds for logic nanoelements which, not relying on charge transport, have the advantage of operating with low energy consumption. Among these elements, three-dimensional (3D) magnetic nanostructures are very promising for achieving high integration density \cite{Gubbiotti2019, sahoo2021, sahoo2018, Parkin2008}. Their potential will materialize only when the underlying spin dynamics is understood \cite{Fischer2020,Streubel2016, Fernandez-Pacheco2017}. Nanotubes (NTs) prepared from ferromagnets represent prototypical 3D nanomagnetic structures \cite{Fernandez-Pacheco2017}. They are extremely versatile as their properties change as a function of, both, their geometrical parameters, namely length, inner and outer radius, \cite{Leblond2004,Escrig2008,Landeros2009} and their axial, helical or vortex-like magnetic configuration \cite{Rueffer2012,Salazar2021}. A curvature-induced magnetochiral field originating from dipole-dipole interaction is expected \cite{Hertel2013} and can induce non-reciprocal spin-wave dispersion relations in case of cylindrical NTs with nanometric radii \cite{Streubel2016,Otalora2016,Otalora2017,Salazar2021,Yang2021,PhysRevB.105.104435}. Previous experimental studies based on microtubes prepared from rolled-up ferromagnetic layers \cite{Balhorn2012,Balhorn2013} have not addressed magnetochiral effects as radii were in the micrometer regime. Nanometric ferromagnetic NTs were investigated recently, however, with hexagonal cross sections \cite{Rueffer2012,Giordano2020,Giordano2021,korber2021,Koerber2022arxiv}. In axially magnetized NTs prepared from magnetically isotropic Ni, a series of spin wave resonances were resolved and classified depending on the number of assumed nodal lines in azimuthal direction \cite{Giordano2020}. The phase distribution across the Ni NTs as explored theoretically in Refs. \cite{Yang2021,Koerber2022arxiv} was not discussed. Koerber et al. \cite{korber2021} studied propagating spin waves along a Ni\textsubscript{80}Fe\textsubscript{20} (permalloy) NT and reported asymmetric spin-wave transport originally predicted for cylindrical NTs. Here, the vortex configuration was induced via a growth-induced magnetic anisotropy and not explicitly by the dipolar interaction relevant for the curvature induced non-reciprocity. For the simulations, the authors considered an ideal hexagonal nanotube. They did not take into account symmetry-breaking aspects like slanted end surfaces \cite{Mehlin2018} and vortex-like segments of opposing chirality \cite{Wyss2017} which occur in real nanotubes.\\
\indent In this Letter, we investigate spin wave modes in NTs with a hexagonal cross section prepared from permalloy (Py) grown by a recently developed plasma-enhanced atomic layer deposition (PEALD) process \cite{Giordano2021,GiordanoPhD}. Combining Brillouin Light Scattering (BLS) microscopy [Fig.~\ref{fig:experiment}(a)] and micromagnetic simulations, we explore the nature of modes occurring in different magnetic configurations with and without helically magnetized segments. Depending on the applied magnetic field we resolve a multitude of spin-wave branches consistent with performed micromagnetic simulations. The latter ones consider the real sample geometry and allow us to relate simulated spin wave modes with helical phase patterns to measured branches. The phase patterns substantiate a curvature-induced magnetochiral effect which was predicted first for circular NTs and then for hexagonal cross-sections \cite{Koerber2022arxiv}. We find an unusual nature of confined modes which so far was restricted to magnets with Dzyaloshinskii-Moriya interaction \cite{Zingsem2019,Ping2021} and puts a new spin on 3D nanomagnonics.
\\
\indent The experiments are based on NTs which are positioned in the gap of a coplanar waveguide (CPW) [Fig.~\ref{fig:experiment}(a)]. By preliminary magnetization dynamics measurements on films and NTs \cite{Giordano2021} we have found that the PEALD-grown Py showed a low damping and was magnetically isotropic. Our NTs therefore exhibit a different effective field compared to Ref. \onlinecite{korber2021}. An rf-current $i_{\rm rf}$ applied to the CPW generates a dynamic magnetic field $h_{\rm rf}$. $h_{\rm rf}$ excites spin precession in the adjacent ferromagnetic NT [Fig.~\ref{fig:experiment}(b)] at the given frequency $f$. Spin-precessional motion is detected via BLS microscopy \cite{Demidov2008IEEE} at room temperature by focusing a monochromatic blue laser on the sample's top surface. The investigated samples consist of a 22 nm-thick Py shell covering a hexagonal GaAs nanowire core and a 5 nm thin spacer layer of Al\textsubscript{2}O\textsubscript{3} used to separate Py and GaAs. In Fig.~\ref{fig:experiment}(c) a transmission electron micrograph of the cross section of a NT from the same batch is shown. For the investigated NT on which we focus here we estimate the effective outer radius to be $r_{\rm o}$ = 80 nm and the inner radius to be $r_{\rm i}$ = 58 nm according to Refs. \onlinecite{Giordano2020,GiordanoPhD} (results on a further NT are presented in the supplementary information \cite{supplement}. An external magnetic field $H$ is applied along the NT axis (\textbf{z}-direction). The position of the BLS laser spot is marked with a blue circle in Fig.~\ref{fig:experiment}(b) comparable with the one of the real laser (about 400 nm). It is positioned in the center of the NT between two nanotroughs which are separated by 560 nm. In our experiments, they are expected to operate as microwave-irradiated emitters of short-wave magnons \cite{Davies2015}. Micromagnetic simulations have been performed using OOMMF \cite{1999oommf} with parameters reported in the supplementary information \cite{supplement} and provided us with eigenfrequencies extracted from power spectral density (PSD) spectra and spin-precessional motion visualized using Mayavi \cite{mayavi}.\\ \indent
In Fig.~\ref{fig:experiment}(d) (Fig.~S2) we show the field-dependent BLS spectra detected for sample NT-s1 (NT-s2) under microwave irradiation for which we varied the frequency between 2.5 and 12.5 GHz. The data were acquired for static fields $\mu_0H$ changing from +90 mT to -90 mT. Colored arrows indicate the specific fields for which spectra are displayed in Fig.~\ref{fig:experiment}(e). Several branches of distinguished eigenmodes are resolved. The Py NTs investigated here show richer spectra compared to the recently studied Ni NTs \cite{Giordano2020}. We attribute this observation to the improved damping parameter $\alpha_{\rm Py}=0.013$ of the PEALD-grown Py \cite{Giordano2021} compared to $\alpha_{\rm Ni}=0.045$ for PEALD-grown Ni \cite{GiordanoPhD}. The spectra are also richer compared to Py disks \cite{Neudecker2006} and Py rings \cite{Podbielski2006}. \\ \indent Nearly linear dependencies of $\frac{df}{dH}$ on $H$ are registered for the main identified resonance modes at $\vert \mu_{0}H \vert$ $\geq$ 42 mT [field regime 3 in Fig.~\ref{fig:experiment}(d)]. Here, we categorize the observed branches in three groups I, II, and III by which we subdivide the frequency regime from 2.5 to 12.5 GHz into three parts. In group II we notice an increase of intensity for certain branches near 8 GHz at fields below $+42~$mT, suggesting a change in the magnetic configuration in field regime 2. Similar behavior is observed for the spectra detected in the regime 2 at negative magnetic fields. In field regime 1 at negative fields, strong resonances (dark red) occur at frequencies down to about 2.5 GHz and up to about 9 GHz. Together with the lack of mirror symmetry with respect to $\mu_{0}H$ = 0 these resonances indicate the irreversible reversal process of the nanotube near $H$ = 0. Our simulations (shown in the supplementary information) indicate that a vortex configuration near zero field is formed consistent with low resonance frequencies. Considering the pronounced intensity of branches near 8 GHz, a large part of the NT exhibits a reversed magnetization at -18 mT when coming from positive $H$. In Ref. \onlinecite{Salazar2021}, the authors predicted small (large) resonance frequencies for non-reciprocal spin waves in a vortex (axial) magnetic configuration of a circular NT at zero (large) magnetic field, in agreement with the frequency variation of branches observed in Fig.~\ref{fig:experiment}(d).\\ \indent The richness of the spectra and the resonant modes in group III at large $H$ in Figs.~\ref{fig:experiment}(d) and S3 suggest that not only azimuthal, but also longitudinal confinement is relevant. It is reasonable to assume that the nanotroughs in the NTs act as either confining boundaries like antidots \cite{Neusser2008} or coherently excited spin-wave emitters \cite{Davies2015} inducing interference patterns \cite{JChen2021}. These considerations motivated us to model the irregularly defined NT segments on which BLS spectroscopy was performed and not an infinitely long NT as was done previously. We hence performed simulations of an (approximately) 560-nm-long hexagonal nanotube with slanted edges \cite{supplement}. These end facets were not parallel and designed such that the finite-size NT had slightly different lengths $L'$ and $L''$ on different sides ($L' > L''$). With the help of the static micromagnetic simulations (shown in Fig.~S4) we attribute the field regions 1, 2, and 3 to three distinctly different configurations of NT magnetization \textbf{M}. Coming from large positive $H$, simulations predict an axial (saturated) state (region 3), a mixed state without (region 2) and with (region 1) a Néel-type domain wall (DW), respectively. In regions 2 and 1, parts of the NTs are in a helical magnetic configuration. In the following we compare experiment and simulations and report the emergence of helical spin-wave modes.
\begin{figure}[t]
	\includegraphics[width=0.48\textwidth]{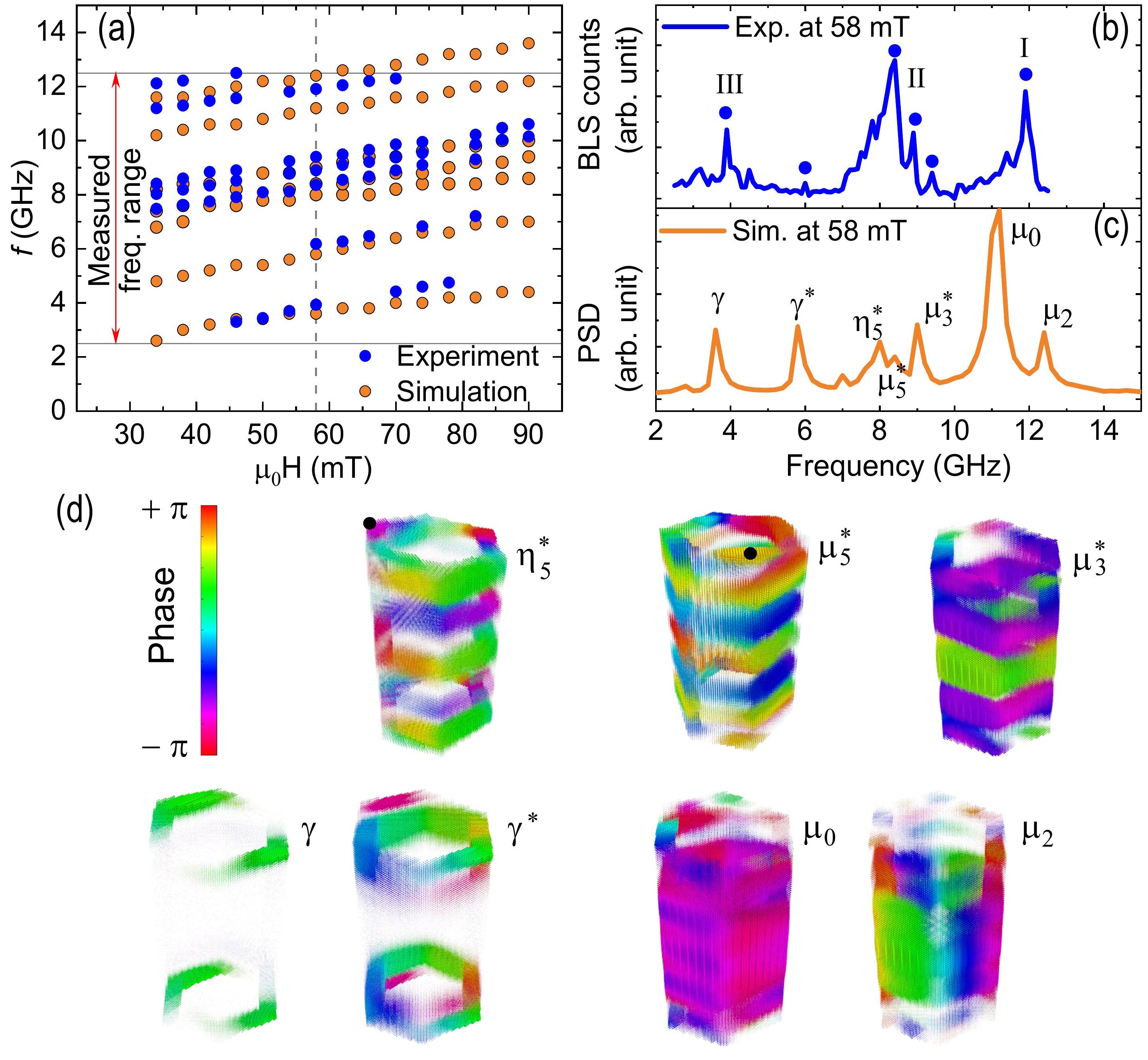}\centering
	\caption{(a) Eigenmode frequencies detected by $\mu$BLS in the central part of a Py NT (blue circles) and simulated resonance frequencies (orange circles) plotted as a function of the magnetic field inducing an axial magnetic state. (b) BLS spectrum and (c) simulated spectrum obtained for a field of 58 mT. Principal eigenmodes are labeled by different letters and correspond to dynamic magnetization profiles in (d). The size of the dots represent the amplitude. The color bar represents spin wave phase ranging from $-\pi$ to $+\pi$. }\label{fig:dynamicSAT}
\end{figure}\\
\indent It is instructive to first discuss the axial state at large positive fields. In Fig.~\ref{fig:dynamicSAT}(a) we summarize resonance frequencies extracted from $\mu$-BLS between nanotroughs of Py NT-s1 (blue circles) and the simulated resonance frequencies (orange circles) plotted as a function of $H$. Experimentally determined resonance frequencies match well with simulated ones (see also Fig.~S5). Representative spectra are displayed for 58 mT in Fig.~\ref{fig:dynamicSAT}(b) and (c) for the experiment and the simulation, respectively. In Fig.~\ref{fig:dynamicSAT}(b) labels I, II, and III are according to the groups of resonances defined in Fig.~\ref{fig:experiment}(d). In Fig.~\ref{fig:dynamicSAT}(c) the Greek symbols refer to the dynamic magnetization profiles (mode patterns) shown in Fig.~\ref{fig:experiment}(d) which are representative for all the fields of the axial (saturated) magnetic state. In the following we discuss the frequency values of resonant modes, and do not refer to the predicted intensities as the selectivity of the BLS microscope concerning specific mode patterns has not been simulated. Simulated modes $\mu_{0}$ and $\mu_{2}$ occurring at 11.2 and 12.4 GHz, respectively, belong to group I. The mode profile $\mu_{0}$ (Fig.~\ref{fig:dynamicSAT}(d) and Supplemental Movie1 \cite{supplement}) corresponds to an in-phase spin precession, which is nearly uniform across the NT and can be considered as the ferromagnetic resonance (FMR) with a total wave vector $k=\sqrt{k^2_z+k^2_\varphi}=0$, where $\mathbf{k}_z$ ($\mathbf{k}_\varphi$) denote wave vectors in $\mathbf{z}$ (azimuthal) direction \cite{Salazar2021}. The mode $\mu_{2}$ illustrates an azimuthal spin wave (Supplemental Movie2 \cite{supplement}) with $k_z=0$ and an azimuthal wave vector $k_\varphi=\nu \times 2\pi / C=26.2$~rad$/\mu$m assuming $\nu=2$. $\nu$ counts the periods in azimuthal ($\varphi$) direction \cite{Koerber2022arxiv}. $C=480~$nm is the circumference of the hexagonal NT as defined in Ref. \onlinecite{Giordano2020}. Considering $\mathbf{k}_\varphi\perp \mathbf{M}$, the mode reflects the Damon-Eshbach (DE) configuration. Consistent with Ref. \cite{Koerber2022arxiv}, its frequency is larger than the FMR. We note that clockwise and counterclockwise azimuthal modes with $\nu \neq 0$ are split in eigenfrequency by the topological Aharonov-Bohm effect and a standing wave is not formed in azimuthal direction \cite{Yang2021,Koerber2022arxiv}.\\
\indent Simulated modes labeled as $\mu^{*}_{3}$, $\mu^{*}_{5}$ and $\eta^{*}_{5}$ have eigenfrequencies of 9, 8.4 and 8 GHz, respectively, consistent with group II resonances. The mode profile of $\mu^{*}_{3}$ agrees with a standing wave confined along a fixed length $L$ with a non-zero $k_z= m\pi/L$ where $m=1,2,..$ and $\nu=0$. In this case, \textbf{k}$_z$ is parallel to \textbf{M} consistent with a backward volume magnetostatic spin wave (BVMSW) configuration. Such modes do not have a pronounced non-reciprocity in axially magnetized NTs \cite{Salazar2021}. Hence, in Fig.~\ref{fig:dynamicSAT}(d) (see Supplemental Movie3 \cite{supplement}), the profile $\mu^{*}_{3}$ incorporates two nearly parallel nodal lines. In Fig.~\ref{fig:dynamicSAT}(d) the mode profiles of $\mu^{*}_{5}$ and $\eta^{*}_{5}$ are rotated to position in each case the facet with the predominant spin-precessional amplitude on the right side (The profile of $\eta^{*}_{5}$ was rotated by 120$^\circ$ anticlockwise around the \textbf{z}-axis with respect to the profile of $\mu^{*}_{5}$). The same corners are marked with black circles. We find that the mode profiles $\mu^{*}_{5}$ and $\eta^{*}_{5}$ represent standing waves with $m=5$ confined along facets of different lengths $L'$ and $L''$, respectively (see Supplemental Movie4 and Movie5 \cite{supplement}). The different longitudinal confinement explains the discrepancy in frequency. The modes labeled $\gamma$ and $\gamma^{*}$ belong to group III with resonant frequencies 3.6 and 5.8 GHz, respectively (see Supplemental Movie6 and Movie7 \cite{supplement}). Here, spin precession occurs right at the NT edges. Such edge modes \cite{Jorzick2002} have the lowest frequencies due to the demagnetizing effect and small internal fields at edges and the nanotroughs \cite{Gurevich96}. The fundamental and first higher order edge mode with larger $k_z$ are separated by about 2.2 GHz which we attribute mainly to exchange interaction. \\ \indent
\begin{figure}[h!]
\includegraphics[width=0.35\textwidth]{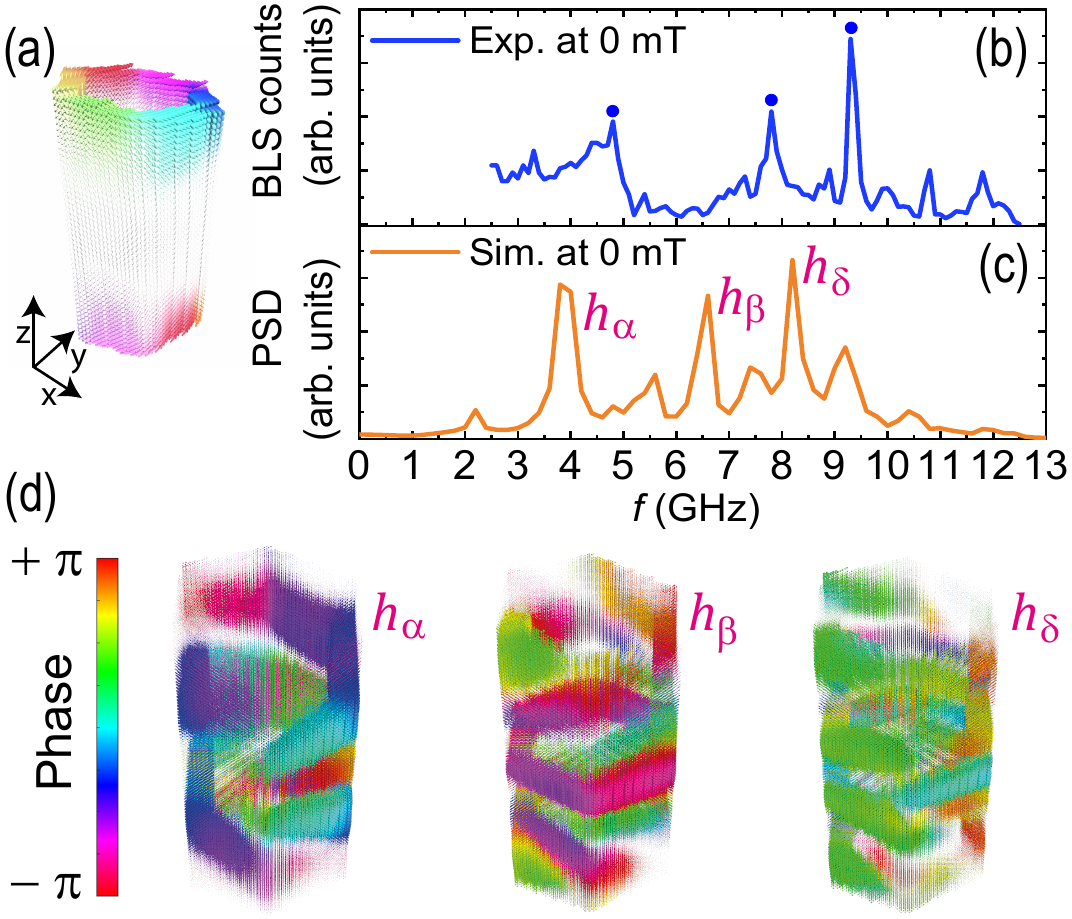}\centering
\caption{(a) Simulated static magnetization profile of the NT at $\mu_{0}H_{z} = 0$~mT. (b) NT-s1 BLS spectrum and (c) simulated spectrum at 0 mT. Specific eigenmodes are labeled by $h_\alpha$, $h_\beta$, and $h_\delta$ and tentatively attributed to modes marked by dots in (b). (d) Dynamic magnetization profiles of modes defined in (c). The size of the dots represent the amplitude. The color bar represents spin wave phase ranging from $-\pi$ to $+\pi$.}\label{fig:dynamicGround}
\end{figure}
As $H$ is reduced, the mixed state is formed with magnetization vortices of opposite chirality at the ends of the NT and neighboring helically aligned segments [Fig.~\ref{fig:dynamicGround}(a)]. Following Ref. \onlinecite{Salazar2021}, NT segments with a helical (vortex) magnetic configuration support non-reciprocal spin waves for non-zero $k$ ($k_z$). The measured (simulated) spectrum at zero field is shown in Fig.~\ref{fig:dynamicGround}(b) [Fig.~\ref{fig:dynamicGround}(c)]. Simulated phase distributions are displayed in Fig. \ref{fig:dynamicGround}(d). Mode $h_{\alpha}$ shown in Fig. \ref{fig:dynamicGround}(d) (Supplemental Movie8 \cite{supplement}) derives most probably from the low-frequency edge mode discussed before. When extending into the helically aligned center region, the mode exhibits a complex helical phase pattern as non-reciprocal spin waves can not form the regular standing wave patterns \cite{Zingsem2019}. Modes $h_{\beta}$ and $h_{\delta}$ (see Supplemental Movie9 and Movie10 \cite{supplement}) reside at higher $f$ suggesting a larger $k_z$. Different from mode $\mu^{*}_{3}$ of the axially aligned state, regular phase patterns with nearly parallel nodal lines are not retrieved in the helical state.\\ \indent Comparing simulated and experimental spectra in Fig.~\ref{fig:dynamicSAT} and Fig.~\ref{fig:dynamicGround}, we observe an overall discrepancy of approximately 1 GHz between prominent peaks in measured (indicated by dots) and simulated (indicated by labels) spectra. We explain the discrepancy with the geometry of the real NT which extends beyond the nanotroughs, leading to an overall smaller demagnetization effect compared to the simulated NT. The latter one has hence a smaller internal field, i.e., smaller eigenfrequencies \cite{Gurevich96} than measured NT.\\ \indent
We now discuss spin waves in the mixed state at $\mu_{0}H_{z} = -14$~mT in which a central DW is assumed [Fig.~\ref{fig:dynamicSwitch}(a)]. The simulation shows two end vortices of opposite chirality which extend to the center and meet in a Néel-type DW (white region). The detected BLS spectrum at -14 mT is reported in Fig.~\ref{fig:dynamicSwitch}(b) and contains at least three groups of peaks near 4, 6.5 and 8.5 GHz. In the simulated spectrum [Fig.~\ref{fig:dynamicSwitch}(c)] also different groups are identified which we label by d, m and v, respectively. In each group a peak is selected (marked with a black tick at the bottom). The corresponding phase pattern and amplitude profile are displayed in Fig.~\ref{fig:dynamicSwitch}(d) and (e), respectively. The amplitudes are extracted along a line in $z$-direction. Vertical dashed lines indicate the width of the DW located near $z=280$~nm. The low frequency resonance at 2 GHz is attributed to a DW resonance (d-mode, Supplemental Movie11 \cite{supplement}). This mode forms a ring of spin precession around the nanotube with weak excitation outside. At 6.6 GHz we find spin-precessional motion with nodes outside the DW (m-mode, Supplemental Movie12 \cite{supplement}). This mode thereby exhibits a non-zero $k_z$ within the vortex region.
The corresponding DE configuration explains the increased frequency. At 8.6 GHz (Supplemental Movie13 \cite{supplement}) we extract six clearly defined nodes along the NT, i.e., $m=7$, giving rise to an even larger $k_z$. The corresponding wavelength amounts to 160 nm and is in the dipole-exchange regime of spin waves. Analyzing the phase profile in Fig.~\ref{fig:dynamicSwitch}(d) we find that nodal lines at 8.6 GHz have again a chiral appearance, consistent with the predicted asymmetric spin-wave dispersion relations. \\ \indent
In summary, we explored spin waves in permalloy nanotubes of finite length using microfocus BLS and micromagnetic simulations. The rich spectra of eigenmodes were attributed to spin waves discretized along both the azimuthal and axial directions of the NT. In the axial magnetic state, nodal lines were found to be mainly straight along the circumference. For helical and vortex-like magnetization orientations, however, nodal lines were distorted which we attribute to the magnetochiral field predicted for nanometric ferromagnetic nanotubes. We hence observe an unusual nature of discretized modes reminiscent of magnets subject to Dzyaloshinskii-Moriya interaction. Our findings pave the way for magnetochiral magnonics which is based on 3D device architectures incorporating segments of nanotubular geometry with different magnetic states.\\ \indent
We thank Didier Bouvet, Anna Fontcuberta i Morral, G{\"o}zde T{\"u}t{\"u}nc{\"u}oglu, and Sho Watanabe for support and SNSF for funding via grants 177550 and 197360. M.C. Giordano and M. Hamdi contributed equally to this work.

\bibliographystyle{apsrev4-2}


\begin{thebibliography}{41}%
\makeatletter
\providecommand \@ifxundefined [1]{%
 \@ifx{#1\undefined}
}%
\providecommand \@ifnum [1]{%
 \ifnum #1\expandafter \@firstoftwo
 \else \expandafter \@secondoftwo
 \fi
}%
\providecommand \@ifx [1]{%
 \ifx #1\expandafter \@firstoftwo
 \else \expandafter \@secondoftwo
 \fi
}%
\providecommand \natexlab [1]{#1}%
\providecommand \enquote  [1]{``#1''}%
\providecommand \bibnamefont  [1]{#1}%
\providecommand \bibfnamefont [1]{#1}%
\providecommand \citenamefont [1]{#1}%
\providecommand \href@noop [0]{\@secondoftwo}%
\providecommand \href [0]{\begingroup \@sanitize@url \@href}%
\providecommand \@href[1]{\@@startlink{#1}\@@href}%
\providecommand \@@href[1]{\endgroup#1\@@endlink}%
\providecommand \@sanitize@url [0]{\catcode `\\12\catcode `\$12\catcode
  `\&12\catcode `\#12\catcode `\^12\catcode `\_12\catcode `\%12\relax}%
\providecommand \@@startlink[1]{}%
\providecommand \@@endlink[0]{}%
\providecommand \url  [0]{\begingroup\@sanitize@url \@url }%
\providecommand \@url [1]{\endgroup\@href {#1}{\urlprefix }}%
\providecommand \urlprefix  [0]{URL }%
\providecommand \Eprint [0]{\href }%
\providecommand \doibase [0]{https://doi.org/}%
\providecommand \selectlanguage [0]{\@gobble}%
\providecommand \bibinfo  [0]{\@secondoftwo}%
\providecommand \bibfield  [0]{\@secondoftwo}%
\providecommand \translation [1]{[#1]}%
\providecommand \BibitemOpen [0]{}%
\providecommand \bibitemStop [0]{}%
\providecommand \bibitemNoStop [0]{.\EOS\space}%
\providecommand \EOS [0]{\spacefactor3000\relax}%
\providecommand \BibitemShut  [1]{\csname bibitem#1\endcsname}%
\let\auto@bib@innerbib\@empty
\bibitem [{\citenamefont {Kruglyak}\ \emph {et~al.}(2010)\citenamefont
  {Kruglyak}, \citenamefont {Demokritov},\ and\ \citenamefont
  {Grundler}}]{Kruglyak2010}%
  \BibitemOpen
  \bibfield  {author} {\bibinfo {author} {\bibfnamefont {V.~V.}\ \bibnamefont
  {Kruglyak}}, \bibinfo {author} {\bibfnamefont {S.~O.}\ \bibnamefont
  {Demokritov}},\ and\ \bibinfo {author} {\bibfnamefont {D.}~\bibnamefont
  {Grundler}},\ }\href {https://doi.org/10.1088/0022-3727/43/26/264001}
  {\bibfield  {journal} {\bibinfo  {journal} {J. Phys. D: Appl. Phys.}\
  }\textbf {\bibinfo {volume} {43}},\ \bibinfo {pages} {264001} (\bibinfo
  {year} {2010})}\BibitemShut {NoStop}%
\bibitem [{\citenamefont {Vogt}\ \emph {et~al.}(2014)\citenamefont {Vogt},
  \citenamefont {Fradin}, \citenamefont {Pearson}, \citenamefont {Sebastian},
  \citenamefont {Bader}, \citenamefont {Hillebrands}, \citenamefont
  {Hoffmann},\ and\ \citenamefont {Schultheiss}}]{Vogt2014}%
  \BibitemOpen
  \bibfield  {author} {\bibinfo {author} {\bibfnamefont {K.}~\bibnamefont
  {Vogt}}, \bibinfo {author} {\bibfnamefont {F.}~\bibnamefont {Fradin}},
  \bibinfo {author} {\bibfnamefont {J.}~\bibnamefont {Pearson}}, \bibinfo
  {author} {\bibfnamefont {T.}~\bibnamefont {Sebastian}}, \bibinfo {author}
  {\bibfnamefont {S.}~\bibnamefont {Bader}}, \bibinfo {author} {\bibfnamefont
  {B.}~\bibnamefont {Hillebrands}}, \bibinfo {author} {\bibfnamefont
  {A.}~\bibnamefont {Hoffmann}},\ and\ \bibinfo {author} {\bibfnamefont
  {H.}~\bibnamefont {Schultheiss}},\ }\href@noop {} {\bibfield  {journal}
  {\bibinfo  {journal} {Nature Communication}\ }\textbf {\bibinfo {volume}
  {5}},\ \bibinfo {pages} {3727} (\bibinfo {year} {2014})}\BibitemShut
  {NoStop}%
\bibitem [{\citenamefont {Gubbiotti}(2019)}]{Gubbiotti2019}%
  \BibitemOpen
  \bibfield  {author} {\bibinfo {author} {\bibfnamefont {G.}~\bibnamefont
  {Gubbiotti}},\ }\href@noop {} {\emph {\bibinfo {title} {Three-Dimensional
  Magnonics: Layered, Micro-and Nanostructures; 1st ed}}}\ (\bibinfo
  {publisher} {Jenny Stanford Publishing.},\ \bibinfo {year}
  {2019})\BibitemShut {NoStop}%
\bibitem [{\citenamefont {Sahoo}\ \emph {et~al.}(2021)\citenamefont {Sahoo},
  \citenamefont {May}, \citenamefont {van Den~Berg}, \citenamefont {Mondal},
  \citenamefont {Ladak},\ and\ \citenamefont {Barman}}]{sahoo2021}%
  \BibitemOpen
  \bibfield  {author} {\bibinfo {author} {\bibfnamefont {S.}~\bibnamefont
  {Sahoo}}, \bibinfo {author} {\bibfnamefont {A.}~\bibnamefont {May}}, \bibinfo
  {author} {\bibfnamefont {A.}~\bibnamefont {van Den~Berg}}, \bibinfo {author}
  {\bibfnamefont {A.~K.}\ \bibnamefont {Mondal}}, \bibinfo {author}
  {\bibfnamefont {S.}~\bibnamefont {Ladak}},\ and\ \bibinfo {author}
  {\bibfnamefont {A.}~\bibnamefont {Barman}},\ }\href
  {https://doi.org/10.1021/acs.nanolett.1c00650} {\bibfield  {journal}
  {\bibinfo  {journal} {Nano Letters}\ }\textbf {\bibinfo {volume} {21}},\
  \bibinfo {pages} {4629} (\bibinfo {year} {2021})}\BibitemShut {NoStop}%
\bibitem [{\citenamefont {Sahoo}\ \emph {et~al.}(2018)\citenamefont {Sahoo},
  \citenamefont {Mondal}, \citenamefont {Williams}, \citenamefont {May},
  \citenamefont {Ladak},\ and\ \citenamefont {Barman}}]{sahoo2018}%
  \BibitemOpen
  \bibfield  {author} {\bibinfo {author} {\bibfnamefont {S.}~\bibnamefont
  {Sahoo}}, \bibinfo {author} {\bibfnamefont {S.}~\bibnamefont {Mondal}},
  \bibinfo {author} {\bibfnamefont {G.}~\bibnamefont {Williams}}, \bibinfo
  {author} {\bibfnamefont {A.}~\bibnamefont {May}}, \bibinfo {author}
  {\bibfnamefont {S.}~\bibnamefont {Ladak}},\ and\ \bibinfo {author}
  {\bibfnamefont {A.}~\bibnamefont {Barman}},\ }\href
  {https://doi.org/10.1039/C7NR07843A} {\bibfield  {journal} {\bibinfo
  {journal} {Nanoscale}\ }\textbf {\bibinfo {volume} {10}},\ \bibinfo {pages}
  {9981} (\bibinfo {year} {2018})}\BibitemShut {NoStop}%
\bibitem [{\citenamefont {Parkin}\ \emph {et~al.}(2008)\citenamefont {Parkin},
  \citenamefont {Hayashi},\ and\ \citenamefont {Thomas}}]{Parkin2008}%
  \BibitemOpen
  \bibfield  {author} {\bibinfo {author} {\bibfnamefont {S.~S.~P.}\
  \bibnamefont {Parkin}}, \bibinfo {author} {\bibfnamefont {M.}~\bibnamefont
  {Hayashi}},\ and\ \bibinfo {author} {\bibfnamefont {L.}~\bibnamefont
  {Thomas}},\ }\href {https://doi.org/10.1126/science.1145799} {\bibfield
  {journal} {\bibinfo  {journal} {Science}\ }\textbf {\bibinfo {volume}
  {320}},\ \bibinfo {pages} {190} (\bibinfo {year} {2008})}\BibitemShut
  {NoStop}%
\bibitem [{\citenamefont {Fischer}\ \emph {et~al.}(2020)\citenamefont
  {Fischer}, \citenamefont {Sanz-Hern{\'{a}}ndez}, \citenamefont {Streubel},\
  and\ \citenamefont {Fern{\'{a}}ndez-Pacheco}}]{Fischer2020}%
  \BibitemOpen
  \bibfield  {author} {\bibinfo {author} {\bibfnamefont {P.}~\bibnamefont
  {Fischer}}, \bibinfo {author} {\bibfnamefont {D.}~\bibnamefont
  {Sanz-Hern{\'{a}}ndez}}, \bibinfo {author} {\bibfnamefont {R.}~\bibnamefont
  {Streubel}},\ and\ \bibinfo {author} {\bibfnamefont {A.}~\bibnamefont
  {Fern{\'{a}}ndez-Pacheco}},\ }\href {https://doi.org/10.1063/1.5134474} {\
  \textbf {\bibinfo {volume} {8}},\ \bibinfo {pages} {10701} (\bibinfo {year}
  {2020})}\BibitemShut {NoStop}%
\bibitem [{\citenamefont {Streubel}\ \emph {et~al.}(2016)\citenamefont
  {Streubel}, \citenamefont {Fischer}, \citenamefont {Kronast}, \citenamefont
  {Kravchuk}, \citenamefont {Sheka}, \citenamefont {Gaididei}, \citenamefont
  {Schmidt},\ and\ \citenamefont {Makarov}}]{Streubel2016}%
  \BibitemOpen
  \bibfield  {author} {\bibinfo {author} {\bibfnamefont {R.}~\bibnamefont
  {Streubel}}, \bibinfo {author} {\bibfnamefont {P.}~\bibnamefont {Fischer}},
  \bibinfo {author} {\bibfnamefont {F.}~\bibnamefont {Kronast}}, \bibinfo
  {author} {\bibfnamefont {V.~P.}\ \bibnamefont {Kravchuk}}, \bibinfo {author}
  {\bibfnamefont {D.~D.}\ \bibnamefont {Sheka}}, \bibinfo {author}
  {\bibfnamefont {Y.}~\bibnamefont {Gaididei}}, \bibinfo {author}
  {\bibfnamefont {O.~G.}\ \bibnamefont {Schmidt}},\ and\ \bibinfo {author}
  {\bibfnamefont {D.}~\bibnamefont {Makarov}},\ }\href
  {https://doi.org/10.1088/0022-3727/49/36/363001} {\bibfield  {journal}
  {\bibinfo  {journal} {J. Phys. D: Appl. Phys.}\ }\textbf {\bibinfo {volume}
  {49}},\ \bibinfo {pages} {363001} (\bibinfo {year} {2016})}\BibitemShut
  {NoStop}%
\bibitem [{\citenamefont {Fern{\'{a}}ndez-Pacheco}\ \emph
  {et~al.}(2017)\citenamefont {Fern{\'{a}}ndez-Pacheco}, \citenamefont
  {Streubel}, \citenamefont {Fruchart}, \citenamefont {Hertel}, \citenamefont
  {Fischer},\ and\ \citenamefont {Cowburn}}]{Fernandez-Pacheco2017}%
  \BibitemOpen
  \bibfield  {author} {\bibinfo {author} {\bibfnamefont {A.}~\bibnamefont
  {Fern{\'{a}}ndez-Pacheco}}, \bibinfo {author} {\bibfnamefont
  {R.}~\bibnamefont {Streubel}}, \bibinfo {author} {\bibfnamefont
  {O.}~\bibnamefont {Fruchart}}, \bibinfo {author} {\bibfnamefont
  {R.}~\bibnamefont {Hertel}}, \bibinfo {author} {\bibfnamefont
  {P.}~\bibnamefont {Fischer}},\ and\ \bibinfo {author} {\bibfnamefont {R.~P.}\
  \bibnamefont {Cowburn}},\ }\href {https://doi.org/10.1038/ncomms15756}
  {\bibfield  {journal} {\bibinfo  {journal} {Nat. Commun.}\ }\textbf {\bibinfo
  {volume} {8}},\ \bibinfo {pages} {15756} (\bibinfo {year}
  {2017})}\BibitemShut {NoStop}%
\bibitem [{\citenamefont {Leblond}\ and\ \citenamefont
  {Veerakumar}(2004)}]{Leblond2004}%
  \BibitemOpen
  \bibfield  {author} {\bibinfo {author} {\bibfnamefont {H.}~\bibnamefont
  {Leblond}}\ and\ \bibinfo {author} {\bibfnamefont {V.}~\bibnamefont
  {Veerakumar}},\ }\href {https://doi.org/10.1103/PhysRevB.70.134413}
  {\bibfield  {journal} {\bibinfo  {journal} {Phys. Rev. B}\ }\textbf {\bibinfo
  {volume} {70}},\ \bibinfo {pages} {134413} (\bibinfo {year}
  {2004})}\BibitemShut {NoStop}%
\bibitem [{\citenamefont {Escrig}\ \emph {et~al.}(2008)\citenamefont {Escrig},
  \citenamefont {Bachmann}, \citenamefont {Jing}, \citenamefont {Daub},
  \citenamefont {Altbir},\ and\ \citenamefont {Nielsch}}]{Escrig2008}%
  \BibitemOpen
  \bibfield  {author} {\bibinfo {author} {\bibfnamefont {J.}~\bibnamefont
  {Escrig}}, \bibinfo {author} {\bibfnamefont {J.}~\bibnamefont {Bachmann}},
  \bibinfo {author} {\bibfnamefont {J.}~\bibnamefont {Jing}}, \bibinfo {author}
  {\bibfnamefont {M.}~\bibnamefont {Daub}}, \bibinfo {author} {\bibfnamefont
  {D.}~\bibnamefont {Altbir}},\ and\ \bibinfo {author} {\bibfnamefont
  {K.}~\bibnamefont {Nielsch}},\ }\href
  {https://doi.org/10.1103/PhysRevB.77.214421} {\bibfield  {journal} {\bibinfo
  {journal} {Phys. Rev. B}\ }\textbf {\bibinfo {volume} {77}},\ \bibinfo
  {pages} {214421} (\bibinfo {year} {2008})}\BibitemShut {NoStop}%
\bibitem [{\citenamefont {Landeros}\ \emph {et~al.}(2009)\citenamefont
  {Landeros}, \citenamefont {Suarez}, \citenamefont {Cuchillo},\ and\
  \citenamefont {Vargas}}]{Landeros2009}%
  \BibitemOpen
  \bibfield  {author} {\bibinfo {author} {\bibfnamefont {P.}~\bibnamefont
  {Landeros}}, \bibinfo {author} {\bibfnamefont {O.~J.}\ \bibnamefont
  {Suarez}}, \bibinfo {author} {\bibfnamefont {A.}~\bibnamefont {Cuchillo}},\
  and\ \bibinfo {author} {\bibfnamefont {P.}~\bibnamefont {Vargas}},\ }\href
  {https://doi.org/10.1103/PhysRevB.79.024404} {\bibfield  {journal} {\bibinfo
  {journal} {Phys. Rev. B}\ }\textbf {\bibinfo {volume} {79}},\ \bibinfo
  {pages} {024404} (\bibinfo {year} {2009})}\BibitemShut {NoStop}%
\bibitem [{\citenamefont {Rüffer}\ \emph {et~al.}(2012)\citenamefont
  {Rüffer}, \citenamefont {Huber}, \citenamefont {Berberich}, \citenamefont
  {Albert}, \citenamefont {Russo-Averchi}, \citenamefont {Heiss}, \citenamefont
  {Arbiol}, \citenamefont {{Fontcuberta i Morral}},\ and\ \citenamefont
  {Grundler}}]{Rueffer2012}%
  \BibitemOpen
  \bibfield  {author} {\bibinfo {author} {\bibfnamefont {D.}~\bibnamefont
  {Rüffer}}, \bibinfo {author} {\bibfnamefont {R.}~\bibnamefont {Huber}},
  \bibinfo {author} {\bibfnamefont {P.}~\bibnamefont {Berberich}}, \bibinfo
  {author} {\bibfnamefont {S.}~\bibnamefont {Albert}}, \bibinfo {author}
  {\bibfnamefont {E.}~\bibnamefont {Russo-Averchi}}, \bibinfo {author}
  {\bibfnamefont {M.}~\bibnamefont {Heiss}}, \bibinfo {author} {\bibfnamefont
  {J.}~\bibnamefont {Arbiol}}, \bibinfo {author} {\bibfnamefont
  {A.}~\bibnamefont {{Fontcuberta i Morral}}},\ and\ \bibinfo {author}
  {\bibfnamefont {D.}~\bibnamefont {Grundler}},\ }\href@noop {} {\bibfield
  {journal} {\bibinfo  {journal} {Nanoscale}\ }\textbf {\bibinfo {volume}
  {4}},\ \bibinfo {pages} {4989} (\bibinfo {year} {2012})}\BibitemShut
  {NoStop}%
\bibitem [{\citenamefont {Salazar-Cardona}\ \emph {et~al.}(2021)\citenamefont
  {Salazar-Cardona}, \citenamefont {Körber}, \citenamefont {Schultheiss},
  \citenamefont {Lenz}, \citenamefont {Thomas}, \citenamefont {Nielsch},
  \citenamefont {Kákay},\ and\ \citenamefont {Otálora}}]{Salazar2021}%
  \BibitemOpen
  \bibfield  {author} {\bibinfo {author} {\bibfnamefont {M.~M.}\ \bibnamefont
  {Salazar-Cardona}}, \bibinfo {author} {\bibfnamefont {L.}~\bibnamefont
  {Körber}}, \bibinfo {author} {\bibfnamefont {H.}~\bibnamefont
  {Schultheiss}}, \bibinfo {author} {\bibfnamefont {K.}~\bibnamefont {Lenz}},
  \bibinfo {author} {\bibfnamefont {A.}~\bibnamefont {Thomas}}, \bibinfo
  {author} {\bibfnamefont {K.}~\bibnamefont {Nielsch}}, \bibinfo {author}
  {\bibfnamefont {A.}~\bibnamefont {Kákay}},\ and\ \bibinfo {author}
  {\bibfnamefont {J.~A.}\ \bibnamefont {Otálora}},\ }\href
  {https://doi.org/10.1063/5.0048692} {\bibfield  {journal} {\bibinfo
  {journal} {Applied Physics Letters}\ }\textbf {\bibinfo {volume} {118}},\
  \bibinfo {pages} {262411} (\bibinfo {year} {2021})}\BibitemShut {NoStop}%
\bibitem [{\citenamefont {Hertel}(2013)}]{Hertel2013}%
  \BibitemOpen
  \bibfield  {author} {\bibinfo {author} {\bibfnamefont {R.}~\bibnamefont
  {Hertel}},\ }\href {https://doi.org/10.1142/S2010324713400092} {\bibfield
  {journal} {\bibinfo  {journal} {SPIN}\ }\textbf {\bibinfo {volume} {03}},\
  \bibinfo {pages} {1340009} (\bibinfo {year} {2013})}\BibitemShut {NoStop}%
\bibitem [{\citenamefont {Ot{\'{a}}lora}\ \emph {et~al.}(2016)\citenamefont
  {Ot{\'{a}}lora}, \citenamefont {Yan}, \citenamefont {Schultheiss},
  \citenamefont {Hertel},\ and\ \citenamefont {K{\'{a}}kay}}]{Otalora2016}%
  \BibitemOpen
  \bibfield  {author} {\bibinfo {author} {\bibfnamefont {J.~A.}\ \bibnamefont
  {Ot{\'{a}}lora}}, \bibinfo {author} {\bibfnamefont {M.}~\bibnamefont {Yan}},
  \bibinfo {author} {\bibfnamefont {H.}~\bibnamefont {Schultheiss}}, \bibinfo
  {author} {\bibfnamefont {R.}~\bibnamefont {Hertel}},\ and\ \bibinfo {author}
  {\bibfnamefont {A.}~\bibnamefont {K{\'{a}}kay}},\ }\href
  {https://doi.org/10.1103/PhysRevLett.117.227203} {\bibfield  {journal}
  {\bibinfo  {journal} {Phys. Rev. Lett.}\ }\textbf {\bibinfo {volume} {117}},\
  \bibinfo {pages} {227203} (\bibinfo {year} {2016})}\BibitemShut {NoStop}%
\bibitem [{\citenamefont {Ot{\'{a}}lora}\ \emph {et~al.}(2017)\citenamefont
  {Ot{\'{a}}lora}, \citenamefont {Yan}, \citenamefont {Schultheiss},
  \citenamefont {Hertel},\ and\ \citenamefont {K{\'{a}}kay}}]{Otalora2017}%
  \BibitemOpen
  \bibfield  {author} {\bibinfo {author} {\bibfnamefont {J.~A.}\ \bibnamefont
  {Ot{\'{a}}lora}}, \bibinfo {author} {\bibfnamefont {M.}~\bibnamefont {Yan}},
  \bibinfo {author} {\bibfnamefont {H.}~\bibnamefont {Schultheiss}}, \bibinfo
  {author} {\bibfnamefont {R.}~\bibnamefont {Hertel}},\ and\ \bibinfo {author}
  {\bibfnamefont {A.}~\bibnamefont {K{\'{a}}kay}},\ }\href
  {https://doi.org/10.1103/PhysRevB.95.184415} {\bibfield  {journal} {\bibinfo
  {journal} {Phys. Rev. B}\ }\textbf {\bibinfo {volume} {95}},\ \bibinfo
  {pages} {184415} (\bibinfo {year} {2017})}\BibitemShut {NoStop}%
\bibitem [{\citenamefont {Yang}\ \emph {et~al.}(2021)\citenamefont {Yang},
  \citenamefont {Yin}, \citenamefont {Li}, \citenamefont {Zeng},\ and\
  \citenamefont {Yan}}]{Yang2021}%
  \BibitemOpen
  \bibfield  {author} {\bibinfo {author} {\bibfnamefont {M.}~\bibnamefont
  {Yang}}, \bibinfo {author} {\bibfnamefont {B.}~\bibnamefont {Yin}}, \bibinfo
  {author} {\bibfnamefont {Z.}~\bibnamefont {Li}}, \bibinfo {author}
  {\bibfnamefont {X.}~\bibnamefont {Zeng}},\ and\ \bibinfo {author}
  {\bibfnamefont {M.}~\bibnamefont {Yan}},\ }\href
  {https://doi.org/10.1103/PhysRevB.103.094404} {\bibfield  {journal} {\bibinfo
   {journal} {Phys. Rev. B}\ }\textbf {\bibinfo {volume} {103}},\ \bibinfo
  {pages} {094404} (\bibinfo {year} {2021})}\BibitemShut {NoStop}%
\bibitem [{\citenamefont {Gallardo}\ \emph {et~al.}(2022)\citenamefont
  {Gallardo}, \citenamefont {Alvarado-Seguel},\ and\ \citenamefont
  {Landeros}}]{PhysRevB.105.104435}%
  \BibitemOpen
  \bibfield  {author} {\bibinfo {author} {\bibfnamefont {R.~A.}\ \bibnamefont
  {Gallardo}}, \bibinfo {author} {\bibfnamefont {P.}~\bibnamefont
  {Alvarado-Seguel}},\ and\ \bibinfo {author} {\bibfnamefont {P.}~\bibnamefont
  {Landeros}},\ }\href {https://doi.org/10.1103/PhysRevB.105.104435} {\bibfield
   {journal} {\bibinfo  {journal} {Phys. Rev. B}\ }\textbf {\bibinfo {volume}
  {105}},\ \bibinfo {pages} {104435} (\bibinfo {year} {2022})}\BibitemShut
  {NoStop}%
\bibitem [{\citenamefont {Balhorn}\ \emph {et~al.}(2012)\citenamefont
  {Balhorn}, \citenamefont {Jeni}, \citenamefont {Hansen}, \citenamefont
  {Heitmann},\ and\ \citenamefont {Mendach}}]{Balhorn2012}%
  \BibitemOpen
  \bibfield  {author} {\bibinfo {author} {\bibfnamefont {F.}~\bibnamefont
  {Balhorn}}, \bibinfo {author} {\bibfnamefont {S.}~\bibnamefont {Jeni}},
  \bibinfo {author} {\bibfnamefont {W.}~\bibnamefont {Hansen}}, \bibinfo
  {author} {\bibfnamefont {D.}~\bibnamefont {Heitmann}},\ and\ \bibinfo
  {author} {\bibfnamefont {S.}~\bibnamefont {Mendach}},\ }\href
  {https://doi.org/10.1063/1.3700809} {\bibfield  {journal} {\bibinfo
  {journal} {Appl. Phys. Lett.}\ }\textbf {\bibinfo {volume} {100}},\ \bibinfo
  {pages} {222402} (\bibinfo {year} {2012})}\BibitemShut {NoStop}%
\bibitem [{\citenamefont {Balhorn}\ \emph {et~al.}(2013)\citenamefont
  {Balhorn}, \citenamefont {Bausch}, \citenamefont {Jeni}, \citenamefont
  {Hansen}, \citenamefont {Heitmann},\ and\ \citenamefont
  {Mendach}}]{Balhorn2013}%
  \BibitemOpen
  \bibfield  {author} {\bibinfo {author} {\bibfnamefont {F.}~\bibnamefont
  {Balhorn}}, \bibinfo {author} {\bibfnamefont {C.}~\bibnamefont {Bausch}},
  \bibinfo {author} {\bibfnamefont {S.}~\bibnamefont {Jeni}}, \bibinfo {author}
  {\bibfnamefont {W.}~\bibnamefont {Hansen}}, \bibinfo {author} {\bibfnamefont
  {D.}~\bibnamefont {Heitmann}},\ and\ \bibinfo {author} {\bibfnamefont
  {S.}~\bibnamefont {Mendach}},\ }\href
  {https://doi.org/10.1103/PhysRevB.88.054402} {\bibfield  {journal} {\bibinfo
  {journal} {Phys. Rev. B}\ }\textbf {\bibinfo {volume} {88}},\ \bibinfo
  {pages} {054402} (\bibinfo {year} {2013})}\BibitemShut {NoStop}%
\bibitem [{\citenamefont {Giordano}\ \emph {et~al.}(2020)\citenamefont
  {Giordano}, \citenamefont {Baumgaertl}, \citenamefont {Escobar~Steinvall},
  \citenamefont {Gay}, \citenamefont {Vuichard}, \citenamefont {{Fontcuberta i
  Morral}},\ and\ \citenamefont {Grundler}}]{Giordano2020}%
  \BibitemOpen
  \bibfield  {author} {\bibinfo {author} {\bibfnamefont {M.~C.}\ \bibnamefont
  {Giordano}}, \bibinfo {author} {\bibfnamefont {K.}~\bibnamefont
  {Baumgaertl}}, \bibinfo {author} {\bibfnamefont {S.}~\bibnamefont
  {Escobar~Steinvall}}, \bibinfo {author} {\bibfnamefont {J.}~\bibnamefont
  {Gay}}, \bibinfo {author} {\bibfnamefont {M.}~\bibnamefont {Vuichard}},
  \bibinfo {author} {\bibfnamefont {A.}~\bibnamefont {{Fontcuberta i
  Morral}}},\ and\ \bibinfo {author} {\bibfnamefont {D.}~\bibnamefont
  {Grundler}},\ }\href {https://doi.org/10.1021/acsami.0c06879} {\bibfield
  {journal} {\bibinfo  {journal} {ACS Applied Materials \& Interfaces}\
  }\textbf {\bibinfo {volume} {12}},\ \bibinfo {pages} {40443} (\bibinfo {year}
  {2020})}\BibitemShut {NoStop}%
\bibitem [{\citenamefont {Giordano}\ \emph {et~al.}(2021)\citenamefont
  {Giordano}, \citenamefont {Escobar~Steinvall}, \citenamefont {Watanabe},
  \citenamefont {Fontcuberta {i}~Morral},\ and\ \citenamefont
  {Grundler}}]{Giordano2021}%
  \BibitemOpen
  \bibfield  {author} {\bibinfo {author} {\bibfnamefont {M.~C.}\ \bibnamefont
  {Giordano}}, \bibinfo {author} {\bibfnamefont {S.}~\bibnamefont
  {Escobar~Steinvall}}, \bibinfo {author} {\bibfnamefont {S.}~\bibnamefont
  {Watanabe}}, \bibinfo {author} {\bibfnamefont {A.}~\bibnamefont {Fontcuberta
  {i}~Morral}},\ and\ \bibinfo {author} {\bibfnamefont {D.}~\bibnamefont
  {Grundler}},\ }\href {https://doi.org/10.1039/d1nr02291a} {\bibfield
  {journal} {\bibinfo  {journal} {Nanoscale}\ }\textbf {\bibinfo {volume}
  {13}},\ \bibinfo {pages} {1351} (\bibinfo {year} {2021})}\BibitemShut
  {NoStop}%
\bibitem [{\citenamefont {K\"orber}\ \emph {et~al.}(2021)\citenamefont
  {K\"orber}, \citenamefont {Zimmermann}, \citenamefont {Wintz}, \citenamefont
  {Finizio}, \citenamefont {Kronseder}, \citenamefont {Bougeard}, \citenamefont
  {Dirnberger}, \citenamefont {Weigand}, \citenamefont {Raabe}, \citenamefont
  {Ot\'alora}, \citenamefont {Schultheiss}, \citenamefont {Josten},
  \citenamefont {Lindner}, \citenamefont {K\'ezsm\'arki}, \citenamefont
  {Back},\ and\ \citenamefont {K\'akay}}]{korber2021}%
  \BibitemOpen
  \bibfield  {author} {\bibinfo {author} {\bibfnamefont {L.}~\bibnamefont
  {K\"orber}}, \bibinfo {author} {\bibfnamefont {M.}~\bibnamefont
  {Zimmermann}}, \bibinfo {author} {\bibfnamefont {S.}~\bibnamefont {Wintz}},
  \bibinfo {author} {\bibfnamefont {S.}~\bibnamefont {Finizio}}, \bibinfo
  {author} {\bibfnamefont {M.}~\bibnamefont {Kronseder}}, \bibinfo {author}
  {\bibfnamefont {D.}~\bibnamefont {Bougeard}}, \bibinfo {author}
  {\bibfnamefont {F.}~\bibnamefont {Dirnberger}}, \bibinfo {author}
  {\bibfnamefont {M.}~\bibnamefont {Weigand}}, \bibinfo {author} {\bibfnamefont
  {J.}~\bibnamefont {Raabe}}, \bibinfo {author} {\bibfnamefont {J.~A.}\
  \bibnamefont {Ot\'alora}}, \bibinfo {author} {\bibfnamefont {H.}~\bibnamefont
  {Schultheiss}}, \bibinfo {author} {\bibfnamefont {E.}~\bibnamefont {Josten}},
  \bibinfo {author} {\bibfnamefont {J.}~\bibnamefont {Lindner}}, \bibinfo
  {author} {\bibfnamefont {I.}~\bibnamefont {K\'ezsm\'arki}}, \bibinfo {author}
  {\bibfnamefont {C.~H.}\ \bibnamefont {Back}},\ and\ \bibinfo {author}
  {\bibfnamefont {A.}~\bibnamefont {K\'akay}},\ }\href
  {https://doi.org/10.1103/PhysRevB.104.184429} {\bibfield  {journal} {\bibinfo
   {journal} {Phys. Rev. B}\ }\textbf {\bibinfo {volume} {104}},\ \bibinfo
  {pages} {184429} (\bibinfo {year} {2021})}\BibitemShut {NoStop}%
\bibitem [{\citenamefont {K{ö}rber}\ \emph {et~al.}(2022)\citenamefont
  {K{ö}rber}, \citenamefont {K{é}zsm{á}rki},\ and\ \citenamefont
  {K{á}kay}}]{Koerber2022arxiv}%
  \BibitemOpen
  \bibfield  {author} {\bibinfo {author} {\bibfnamefont {L.}~\bibnamefont
  {K{ö}rber}}, \bibinfo {author} {\bibfnamefont {I.}~\bibnamefont
  {K{é}zsm{á}rki}},\ and\ \bibinfo {author} {\bibfnamefont {A.}~\bibnamefont
  {K{á}kay}}} (\bibinfo {year} {2022}),\ \bibinfo {note}
  {arXiv:2202.06601v1}\BibitemShut {NoStop}%
\bibitem [{\citenamefont {Mehlin}\ \emph {et~al.}(lack)\citenamefont {Mehlin},
  \citenamefont {Gross}, \citenamefont {Wyss}, \citenamefont {Schefer},
  \citenamefont {T\"ut\"unc\"uoglu}, \citenamefont {Heimbach}, \citenamefont
  {Fontcuberta~i Morral}, \citenamefont {Grundler},\ and\ \citenamefont
  {Poggio}}]{Mehlin2018}%
  \BibitemOpen
  \bibfield  {author} {\bibinfo {author} {\bibfnamefont {A.}~\bibnamefont
  {Mehlin}}, \bibinfo {author} {\bibfnamefont {B.}~\bibnamefont {Gross}},
  \bibinfo {author} {\bibfnamefont {M.}~\bibnamefont {Wyss}}, \bibinfo {author}
  {\bibfnamefont {T.}~\bibnamefont {Schefer}}, \bibinfo {author} {\bibfnamefont
  {G.}~\bibnamefont {T\"ut\"unc\"uoglu}}, \bibinfo {author} {\bibfnamefont
  {F.}~\bibnamefont {Heimbach}}, \bibinfo {author} {\bibfnamefont
  {A.}~\bibnamefont {Fontcuberta~i Morral}}, \bibinfo {author} {\bibfnamefont
  {D.}~\bibnamefont {Grundler}},\ and\ \bibinfo {author} {\bibfnamefont
  {M.}~\bibnamefont {Poggio}},\ }\href
  {https://doi.org/10.1103/PhysRevB.97.134422} {\bibfield  {journal} {\bibinfo
  {journal} {Phys. Rev. B}\ }\textbf {\bibinfo {volume} {97}},\ \bibinfo
  {pages} {134422} (\bibinfo {year} {2018\color{black}})}\BibitemShut {NoStop}%
\bibitem [{\citenamefont {Wyss}\ \emph {et~al.}(2017)\citenamefont {Wyss},
  \citenamefont {Mehlin}, \citenamefont {Gross}, \citenamefont {Buchter},
  \citenamefont {Farhan}, \citenamefont {Buzzi}, \citenamefont {Kleibert},
  \citenamefont {T\"ut\"unc\"uoglu}, \citenamefont {Heimbach}, \citenamefont
  {Fontcuberta~i Morral}, \citenamefont {Grundler},\ and\ \citenamefont
  {Poggio}}]{Wyss2017}%
  \BibitemOpen
  \bibfield  {author} {\bibinfo {author} {\bibfnamefont {M.}~\bibnamefont
  {Wyss}}, \bibinfo {author} {\bibfnamefont {A.}~\bibnamefont {Mehlin}},
  \bibinfo {author} {\bibfnamefont {B.}~\bibnamefont {Gross}}, \bibinfo
  {author} {\bibfnamefont {A.}~\bibnamefont {Buchter}}, \bibinfo {author}
  {\bibfnamefont {A.}~\bibnamefont {Farhan}}, \bibinfo {author} {\bibfnamefont
  {M.}~\bibnamefont {Buzzi}}, \bibinfo {author} {\bibfnamefont
  {A.}~\bibnamefont {Kleibert}}, \bibinfo {author} {\bibfnamefont
  {G.}~\bibnamefont {T\"ut\"unc\"uoglu}}, \bibinfo {author} {\bibfnamefont
  {F.}~\bibnamefont {Heimbach}}, \bibinfo {author} {\bibfnamefont
  {A.}~\bibnamefont {Fontcuberta~i Morral}}, \bibinfo {author} {\bibfnamefont
  {D.}~\bibnamefont {Grundler}},\ and\ \bibinfo {author} {\bibfnamefont
  {M.}~\bibnamefont {Poggio}},\ }\href
  {https://doi.org/10.1103/PhysRevB.96.024423} {\bibfield  {journal} {\bibinfo
  {journal} {Phys. Rev. B}\ }\textbf {\bibinfo {volume} {96}},\ \bibinfo
  {pages} {024423} (\bibinfo {year} {2017})}\BibitemShut {NoStop}%
\bibitem [{\citenamefont {Giordano}(2021)}]{GiordanoPhD}%
  \BibitemOpen
  \bibfield  {author} {\bibinfo {author} {\bibfnamefont {M.~C.}\ \bibnamefont
  {Giordano}},\ }\emph {\bibinfo {title} {Atomic layer deposition of {N}i and
  {N}i$_{80}${F}e$_{20}$ for tubular spin-wave nanocavities}},\ \href@noop {}
  {\bibinfo {type} {Ph{D} thesis}},\ \bibinfo  {school} {EPFL} (\bibinfo {year}
  {2021})\BibitemShut {NoStop}%
\bibitem [{\citenamefont {Zingsem}\ \emph {et~al.}(2019)\citenamefont
  {Zingsem}, \citenamefont {Farle}, \citenamefont {Stamps},\ and\ \citenamefont
  {Camley}}]{Zingsem2019}%
  \BibitemOpen
  \bibfield  {author} {\bibinfo {author} {\bibfnamefont {B.~W.}\ \bibnamefont
  {Zingsem}}, \bibinfo {author} {\bibfnamefont {M.}~\bibnamefont {Farle}},
  \bibinfo {author} {\bibfnamefont {R.~L.}\ \bibnamefont {Stamps}},\ and\
  \bibinfo {author} {\bibfnamefont {R.~E.}\ \bibnamefont {Camley}},\ }\href
  {https://doi.org/10.1103/PhysRevB.99.214429} {\bibfield  {journal} {\bibinfo
  {journal} {Phys. Rev. B}\ }\textbf {\bibinfo {volume} {99}},\ \bibinfo
  {pages} {214429} (\bibinfo {year} {2019})}\BibitemShut {NoStop}%
\bibitem [{\citenamefont {Che}\ \emph {et~al.}(2021)\citenamefont {Che},
  \citenamefont {Stasinopoulos}, \citenamefont {Mucchietto}, \citenamefont
  {Li}, \citenamefont {Berger}, \citenamefont {Bauer}, \citenamefont
  {Pfleiderer},\ and\ \citenamefont {Grundler}}]{Ping2021}%
  \BibitemOpen
  \bibfield  {author} {\bibinfo {author} {\bibfnamefont {P.}~\bibnamefont
  {Che}}, \bibinfo {author} {\bibfnamefont {I.}~\bibnamefont {Stasinopoulos}},
  \bibinfo {author} {\bibfnamefont {A.}~\bibnamefont {Mucchietto}}, \bibinfo
  {author} {\bibfnamefont {J.}~\bibnamefont {Li}}, \bibinfo {author}
  {\bibfnamefont {H.}~\bibnamefont {Berger}}, \bibinfo {author} {\bibfnamefont
  {A.}~\bibnamefont {Bauer}}, \bibinfo {author} {\bibfnamefont
  {C.}~\bibnamefont {Pfleiderer}},\ and\ \bibinfo {author} {\bibfnamefont
  {D.}~\bibnamefont {Grundler}},\ }\href
  {https://doi.org/10.1103/PhysRevResearch.3.033104} {\bibfield  {journal}
  {\bibinfo  {journal} {Phys. Rev. Research}\ }\textbf {\bibinfo {volume}
  {3}},\ \bibinfo {pages} {033104} (\bibinfo {year} {2021})}\BibitemShut
  {NoStop}%
\bibitem [{\citenamefont {Demokritov}\ and\ \citenamefont
  {Demidov}(2008)}]{Demidov2008IEEE}%
  \BibitemOpen
  \bibfield  {author} {\bibinfo {author} {\bibfnamefont {S.~O.}\ \bibnamefont
  {Demokritov}}\ and\ \bibinfo {author} {\bibfnamefont {V.~E.}\ \bibnamefont
  {Demidov}},\ }\href {https://doi.org/10.1109/TMAG.2007.910227} {\bibfield
  {journal} {\bibinfo  {journal} {IEEE Trans. Magn.}\ }\textbf {\bibinfo
  {volume} {44}},\ \bibinfo {pages} {6–12} (\bibinfo {year}
  {2008})}\BibitemShut {NoStop}%
\bibitem [{sup()}]{supplement}%
  \BibitemOpen
  \href@noop {} {}\bibinfo {note} {See Supplemental Material. In the
  supplemental movies (a) represent the spin wave profile across a cross
  sectional cut at the represented height of the NT and (b) shows the spin wave
  profile on the entire NT while NT is rotatig along its axis. The spin wave
  amplitude is represented by size of the dots. The color bar represents spin
  wave phase ranging from $-\pi$ to $+\pi$. Supplemental movies are prepared
  using Mayavi \cite{mayavi}\color{black}}\BibitemShut {NoStop}%
\bibitem [{\citenamefont {Davies}\ \emph {et~al.}(2015)\citenamefont {Davies},
  \citenamefont {Sadovnikov}, \citenamefont {Grishin}, \citenamefont
  {Sharaevskii}, \citenamefont {Nikitov},\ and\ \citenamefont
  {Kruglyak}}]{Davies2015}%
  \BibitemOpen
  \bibfield  {author} {\bibinfo {author} {\bibfnamefont {C.~S.}\ \bibnamefont
  {Davies}}, \bibinfo {author} {\bibfnamefont {A.~V.}\ \bibnamefont
  {Sadovnikov}}, \bibinfo {author} {\bibfnamefont {S.~V.}\ \bibnamefont
  {Grishin}}, \bibinfo {author} {\bibfnamefont {Y.~P.}\ \bibnamefont
  {Sharaevskii}}, \bibinfo {author} {\bibfnamefont {S.~A.}\ \bibnamefont
  {Nikitov}},\ and\ \bibinfo {author} {\bibfnamefont {V.~V.}\ \bibnamefont
  {Kruglyak}},\ }\href {https://doi.org/10.1063/1.4933263} {\bibfield
  {journal} {\bibinfo  {journal} {Appl. Phys. Lett.}\ }\textbf {\bibinfo
  {volume} {107}},\ \bibinfo {pages} {162401} (\bibinfo {year} {2015})},\
  \Eprint {https://arxiv.org/abs/https://doi.org/10.1063/1.4933263}
  {https://doi.org/10.1063/1.4933263} \BibitemShut {NoStop}%
\bibitem [{\citenamefont {Donahue}\ and\ \citenamefont
  {Porter}(1999)}]{1999oommf}%
  \BibitemOpen
  \bibfield  {author} {\bibinfo {author} {\bibfnamefont {M.~J.}\ \bibnamefont
  {Donahue}}\ and\ \bibinfo {author} {\bibfnamefont {D.~G.}\ \bibnamefont
  {Porter}},\ }\href@noop {} {\emph {\bibinfo {title} {OOMMF user's guide,
  version 1.0}}}\ (\bibinfo  {publisher} {US Department of Commerce, National
  Institute of Standards and Technology},\ \bibinfo {year} {1999})\BibitemShut
  {NoStop}%
\bibitem [{\citenamefont {Ramachandran}\ and\ \citenamefont
  {Varoquaux}(2011)}]{mayavi}%
  \BibitemOpen
  \bibfield  {author} {\bibinfo {author} {\bibfnamefont {P.}~\bibnamefont
  {Ramachandran}}\ and\ \bibinfo {author} {\bibfnamefont {G.}~\bibnamefont
  {Varoquaux}},\ }\href@noop {} {\bibfield  {journal} {\bibinfo  {journal}
  {Computing in Science \& Engineering}\ }\textbf {\bibinfo {volume} {13}},\
  \bibinfo {pages} {40} (\bibinfo {year} {2011})},\ \bibinfo {note}
  {\textit{Mayavi: {3D} visualization of scientific
  data}\color{black}}\BibitemShut {NoStop}%
\bibitem [{\citenamefont {Neudecker}\ \emph {et~al.}(2006)\citenamefont
  {Neudecker}, \citenamefont {Perzlmaier}, \citenamefont {Hoffmann},
  \citenamefont {Woltersdorf}, \citenamefont {Buess}, \citenamefont {Weiss},\
  and\ \citenamefont {Back}}]{Neudecker2006}%
  \BibitemOpen
  \bibfield  {author} {\bibinfo {author} {\bibfnamefont {I.}~\bibnamefont
  {Neudecker}}, \bibinfo {author} {\bibfnamefont {K.}~\bibnamefont
  {Perzlmaier}}, \bibinfo {author} {\bibfnamefont {F.}~\bibnamefont
  {Hoffmann}}, \bibinfo {author} {\bibfnamefont {G.}~\bibnamefont
  {Woltersdorf}}, \bibinfo {author} {\bibfnamefont {M.}~\bibnamefont {Buess}},
  \bibinfo {author} {\bibfnamefont {D.}~\bibnamefont {Weiss}},\ and\ \bibinfo
  {author} {\bibfnamefont {C.~H.}\ \bibnamefont {Back}},\ }\href
  {https://doi.org/10.1103/PhysRevB.73.134426} {\bibfield  {journal} {\bibinfo
  {journal} {Phys. Rev. B}\ }\textbf {\bibinfo {volume} {73}},\ \bibinfo
  {pages} {134426} (\bibinfo {year} {2006})}\BibitemShut {NoStop}%
\bibitem [{\citenamefont {Podbielski}\ \emph {et~al.}(2006)\citenamefont
  {Podbielski}, \citenamefont {Giesen},\ and\ \citenamefont
  {Grundler}}]{Podbielski2006}%
  \BibitemOpen
  \bibfield  {author} {\bibinfo {author} {\bibfnamefont {J.}~\bibnamefont
  {Podbielski}}, \bibinfo {author} {\bibfnamefont {F.}~\bibnamefont {Giesen}},\
  and\ \bibinfo {author} {\bibfnamefont {D.}~\bibnamefont {Grundler}},\ }\href
  {https://doi.org/10.1103/PhysRevLett.96.167207} {\bibfield  {journal}
  {\bibinfo  {journal} {Phys. Rev. Lett.}\ }\textbf {\bibinfo {volume} {96}},\
  \bibinfo {pages} {167207} (\bibinfo {year} {2006})}\BibitemShut {NoStop}%
\bibitem [{\citenamefont {Neusser}\ \emph {et~al.}(2008)\citenamefont
  {Neusser}, \citenamefont {Botters},\ and\ \citenamefont
  {Grundler}}]{Neusser2008}%
  \BibitemOpen
  \bibfield  {author} {\bibinfo {author} {\bibfnamefont {S.}~\bibnamefont
  {Neusser}}, \bibinfo {author} {\bibfnamefont {B.}~\bibnamefont {Botters}},\
  and\ \bibinfo {author} {\bibfnamefont {D.}~\bibnamefont {Grundler}},\ }\href
  {https://doi.org/10.1103/PhysRevB.78.054406} {\bibfield  {journal} {\bibinfo
  {journal} {Phys. Rev. B}\ }\textbf {\bibinfo {volume} {78}},\ \bibinfo
  {pages} {054406} (\bibinfo {year} {2008})}\BibitemShut {NoStop}%
\bibitem [{\citenamefont {Chen}\ \emph {et~al.}(2021)\citenamefont {Chen},
  \citenamefont {Wang}, \citenamefont {Hula}, \citenamefont {Liu},
  \citenamefont {Liu}, \citenamefont {Liu}, \citenamefont {Jia}, \citenamefont
  {Song}, \citenamefont {Guo}, \citenamefont {Zhang}, \citenamefont {Zhang},
  \citenamefont {Han}, \citenamefont {Yu}, \citenamefont {Wu}, \citenamefont
  {Schultheiss},\ and\ \citenamefont {Yu}}]{JChen2021}%
  \BibitemOpen
  \bibfield  {author} {\bibinfo {author} {\bibfnamefont {J.}~\bibnamefont
  {Chen}}, \bibinfo {author} {\bibfnamefont {H.}~\bibnamefont {Wang}}, \bibinfo
  {author} {\bibfnamefont {T.}~\bibnamefont {Hula}}, \bibinfo {author}
  {\bibfnamefont {C.}~\bibnamefont {Liu}}, \bibinfo {author} {\bibfnamefont
  {S.}~\bibnamefont {Liu}}, \bibinfo {author} {\bibfnamefont {T.}~\bibnamefont
  {Liu}}, \bibinfo {author} {\bibfnamefont {H.}~\bibnamefont {Jia}}, \bibinfo
  {author} {\bibfnamefont {Q.}~\bibnamefont {Song}}, \bibinfo {author}
  {\bibfnamefont {C.}~\bibnamefont {Guo}}, \bibinfo {author} {\bibfnamefont
  {Y.}~\bibnamefont {Zhang}}, \bibinfo {author} {\bibfnamefont {J.~Z.~J.}\
  \bibnamefont {Zhang}}, \bibinfo {author} {\bibfnamefont {X.}~\bibnamefont
  {Han}}, \bibinfo {author} {\bibfnamefont {D.}~\bibnamefont {Yu}}, \bibinfo
  {author} {\bibfnamefont {M.}~\bibnamefont {Wu}}, \bibinfo {author}
  {\bibfnamefont {H.}~\bibnamefont {Schultheiss}},\ and\ \bibinfo {author}
  {\bibfnamefont {H.}~\bibnamefont {Yu}},\ }\href
  {https://doi.org/10.1021/acs.nanolett.1c02010} {\bibfield  {journal}
  {\bibinfo  {journal} {Nano Lett.}\ }\textbf {\bibinfo {volume} {21}},\
  \bibinfo {pages} {6237} (\bibinfo {year} {2021})}\BibitemShut {NoStop}%
\bibitem [{\citenamefont {Jorzick}\ \emph {et~al.}(2002)\citenamefont
  {Jorzick}, \citenamefont {Demokritov}, \citenamefont {Hillebrands},
  \citenamefont {Bailleul}, \citenamefont {Fermon}, \citenamefont {Guslienko},
  \citenamefont {Slavin}, \citenamefont {Berkov},\ and\ \citenamefont
  {Gorn}}]{Jorzick2002}%
  \BibitemOpen
  \bibfield  {author} {\bibinfo {author} {\bibfnamefont {J.}~\bibnamefont
  {Jorzick}}, \bibinfo {author} {\bibfnamefont {S.~O.}\ \bibnamefont
  {Demokritov}}, \bibinfo {author} {\bibfnamefont {B.}~\bibnamefont
  {Hillebrands}}, \bibinfo {author} {\bibfnamefont {M.}~\bibnamefont
  {Bailleul}}, \bibinfo {author} {\bibfnamefont {C.}~\bibnamefont {Fermon}},
  \bibinfo {author} {\bibfnamefont {K.~Y.}\ \bibnamefont {Guslienko}}, \bibinfo
  {author} {\bibfnamefont {A.~N.}\ \bibnamefont {Slavin}}, \bibinfo {author}
  {\bibfnamefont {D.~V.}\ \bibnamefont {Berkov}},\ and\ \bibinfo {author}
  {\bibfnamefont {N.~L.}\ \bibnamefont {Gorn}},\ }\href
  {https://doi.org/10.1103/PhysRevLett.88.047204} {\bibfield  {journal}
  {\bibinfo  {journal} {Phys. Rev. Lett.}\ }\textbf {\bibinfo {volume} {88}},\
  \bibinfo {pages} {047204} (\bibinfo {year} {2002})}\BibitemShut {NoStop}%
\bibitem [{\citenamefont {Gurevich}\ and\ \citenamefont
  {Melkov}(1996)}]{Gurevich96}%
  \BibitemOpen
  \bibfield  {author} {\bibinfo {author} {\bibfnamefont {A.~G.}\ \bibnamefont
  {Gurevich}}\ and\ \bibinfo {author} {\bibfnamefont {G.~A.}\ \bibnamefont
  {Melkov}},\ }\href
  {https://www.google.com/books?hl=en&lr=&id=YgQtSvFIvFQC&oi=fnd&pg=IA2&dq=Gurevich+1996+Magnetization+oscillations+and+waves+book&ots=PvqcHDuhSc&sig=JUwpbw7MaC9m0a8_3_OSJGuYmYw}
  {\emph {\bibinfo {title} {Magnetization oscillations and waves}}}\ (\bibinfo
  {publisher} {CRC Press, Boca Raton},\ \bibinfo {year} {1996})\BibitemShut
  {NoStop}%
\end{thebibliography}
%

\end{document}